\documentclass[aps,twocolumn,showpacs,superscriptaddress,pra]{revtex4-1}
\usepackage{graphicx}
\graphicspath{{figures_coherence/}{./}}
\usepackage{caption}
\usepackage{subcaption}
\RequirePackage[font=footnotesize,labelfont={bf,it}]{caption}
\usepackage[T1]{fontenc}
\usepackage{textcomp}
\usepackage{amsmath,amssymb}
\usepackage{tikz}
\usepackage{soul}
\usepackage{color}
\usetikzlibrary{positioning,arrows}
\usetikzlibrary{decorations.pathmorphing}
\usetikzlibrary{decorations.markings}
\definecolor{myblue}{RGB}{56,94,141}

\newcommand{\newc}{\newcommand}

\newc{\kt}{\rangle}
\newc{\br}{\langle}

\newc{\pr}{\prime}
\newc{\longra}{\longrightarrow}
\newc{\ot}{\otimes}
\newc{\rarrow}{\rightarrow}
\newc{\h}{\hat}
\newc{\bom}{\boldmath}
\newc{\btd}{\bigtriangledown}
\newc{\al}{\alpha}
\newc{\be}{\beta}
\newc{\ld}{\lambda}
\newc{\sg}{\sigma}
\newc{\p}{\psi}
\newc{\eps}{\epsilon}
\newc{\om}{\omega}
\newc{\mb}{\mbox}
\newc{\tm}{\times}
\newc{\hu}{\hat{u}}
\newc{\hv}{\hat{v}}
\newc{\hk}{\hat{K}}
\newc{\ra}{\rightarrow}
\newc{\non}{\nonumber}

\newc{\dg}{\dagger}
\newc{\prh}{\mbox{PR}_H}
\newc{\prq}{\mbox{PR}_q}
\newc{\tr}{\mbox{tr}}
\newc{\pd}{\partial}
\newc{\qv}{\vec{q}}
\newc{\pv}{\vec{p}}
\newc{\dqv}{\delta\vec{q}}
\newc{\dpv}{\delta\vec{p}}
\newc{\mbq}{\mathbf{q}}
\newc{\mbqp}{\mathbf{q'}}
\newc{\mbpp}{\mathbf{p'}}
\newc{\mbp}{\mathbf{p}}
\newc{\mbn}{\mathbf{\nabla}}
\newc{\dmbq}{\delta \mbq}
\newc{\dmbp}{\delta \mbp}
\newc{\T}{\mathsf{T}}
\newc{\J}{\mathsf{J}}
\newc{\sfL}{\mathsf{L}}
\newc{\C}{\mathsf{C}}
\newc{\B}{\mathsf{M}}
\newc{\V}{\mathsf{V}}
\usepackage{mdframed}
\newcommand{\diag}{\mathop{\rm diag}}

\newcommand{\homega}{\bar \omega}

\newcommand{\bra}[1]{\left\langle #1 \right\vert}
\newcommand{\ket}[1]{\left\vert #1 \right\rangle}

\usepackage{xr}
\usepackage{xr-hyper}

\begin{document}
\tikzset{gluon/.style={decorate,draw=red,decoration={coil,aspect=0}},
photon/.style={decorate,decoration={snake=bumps},draw=blue}
bracing/.style={decorate,decoration=brace,draw=myblue}
crossing/.style= {decorate,decoration=zigzag,pre=crosses,pre length=0.5cm,draw=myblue}
}

\title{Coherence and mixedness of neutrino oscillations in a magnetic field}
\author{P. Kurashvili}
\affiliation{National Centre for Nuclear Research, Warsaw 00-681, Poland}
\author{L. Chotorlishvili}
\affiliation{Institute f\"ur Physik, Martin-Luther Universit\"at Halle-Wittenberg, D-06120 Halle/Saale, Germany}
\author{K. A. Kouzakov}
\affiliation{Department of Nuclear Physics and Quantum Theory of Collisions,
Faculty of Physics, Lomonosov Moscow State University, Moscow 119991, Russia}
\author{A. I. Studenikin}
\affiliation{Department of Theoretical Physics, Faculty of Physics,
Lomonosov Moscow State University, Moscow 119991, Russia\\
and Joint Institute for Nuclear Research,
Dubna 141980, Moscow Region, Russia}

\date{\today}

%
\begin{abstract}
%
The radical departure from classical physics implies quantum coherence, i.e.,  coherent superposition of eigenstates of Hermitian operators with a discrete spectrum. In resource theory, quantum coherence is a resource for quantum operations. Typically the stochastic phenomenon induces decoherence effects. However, in the present work, we prove that nonunitary evolution leads to the generation of quantum coherence in some cases. Specifically, we consider the neutrino propagation in the dissipative environment, namely in a magnetic field with a stochastic component, and focus on neutrino flavor, spin and spin-flavor oscillations. We present exact analytical results for quantum coherence in neutrino oscillations quantified in terms of the relative entropy. Starting from an initial zero coherence state, we observe persistent oscillations of coherence during the dissipative evolution. We found that after dissipative evolution, the initial spin-polarized state entirely thermalizes, and in the final steady state, the spin-up/down states have the same probabilities. On the other hand, neutrino flavor states also thermalize, byt the populations of two flavor states do not equate to each other. The initial flavor still dominates in the final steady state.  
\end{abstract}

\maketitle

\section{Introduction}
Coherence is a hardwired feature of quantum systems, a key ingredient in versatile applications. The resource theory of quantum coherence~\cite{Aaberg,Baumgratz,Siddiqui,Cianciaruso,Susana,Streltsov,Oppenheim,Hai-Qing,Chitambar,Xueyuan} exploits ideas of optimal consumption of resources and proper management of costs. The free states and free operations require per se zero costs.  However, free resources are not enough for quantum information protocols. Rather briefly, we recall the underlying formalism of resource theory.

Let $\mathcal{\hat{A}}$ be the set of free states and $\mathcal{\hat{Q}}$ the set of free quantum operations. In the resource theory one always implies that the following criteria hold: $\mathcal{\hat{Q}}_n\left( \mathcal{\hat{A}}_n\right)\in\mathcal{\hat{A}}$ for $\forall \mathcal{\hat{A}}_n\in \mathcal{\hat{A}}$ and $\forall \mathcal{\hat{Q}}_n\in \mathcal{\hat{Q}}$.  In the entanglement theory, typically $\mathcal{\hat{A}}$ is the set of separable states and $\mathcal{\hat{Q}}$ are local operations and classical communications. 
In the realistic physical systems operations, $\mathcal{\hat{Q}}$ can be provided at a low cost but not for free. Nevertheless, for the open quantum systems, we do not count resources supplied from the environment. 

Up to date, mainly non-relativistic quantum systems were in the scope of the quantum resource theory. However, its concepts are universal and firmly can be extended to the relativistic quantum systems and neutrinos in particular~\cite{dixit2019epjc}. Neutrinos host dichotomic left-right helicity and different lepton flavor (electron, muon, or tau) and, when propagating, they can change their type, or oscillate. Neutrino oscillations is an inherently quantum mechanical phenomenon~\cite{song2018quantifying,PhysRevLett.117.050402} and can be interpreted in terms of quantum resource theory. An interesting case of this phenomenon is expected when neutrinos propagate in the presence of a magnetic field: neutrinos can change both their flavor and helicity (see, for instance, Refs.~\cite{kurashvili2017spin,popov2019epjc} and references therein). The indicated oscillations can serve as a manifestation of new physics, namely neutrino electromagnetic interactions~\cite{rmp2015,andp2016}, and can be especially relevant for cosmic neutrinos that propagate in various astrophysical environments, where nonzero magnetic fields are known to exist.

In the present work, we propose the basis-dependent rigorous formulation of resource theory of coherence for neutrino flavor, spin and spin-flavor oscillations. We exploit the incoherent states $\mathcal{\hat{I}}_n$ as a free state, and a magnetic field we exploit as a source of operations for generating the coherence $\mathcal{\hat{Q}}_n\left( \mathcal{\hat{I}}_n\right)\notin\mathcal{\hat{I}}$. We limit ourselves to two neutrino generations and start with Dirac neutrino helicity basis states $\vert\nu_{1,s=\pm 1}\rangle$, $\vert\nu_{2,s=\pm 1}\rangle$ with masses $m_1$ and $m_2$ ($m_2>m_1$). The neutrino left- and right-handed flavor states are then given by
\begin{eqnarray}\label{flavor basis}
\vert\nu_{e}^{R,L}\rangle&=&\cos\theta\vert\nu_{1,s=\pm 1}\rangle+\sin\theta\vert\nu_{2,s=\pm 1}\rangle,\nonumber\\
\vert\nu_{\mu}^{R,L}\rangle&=&\sin\theta\vert\nu_{1,s=\pm 1}\rangle+\cos\theta\vert\nu_{2,s=\pm 1}\rangle,
\end{eqnarray}
where $\theta$ is the mixing angle and the subscripts $e$ and $\mu$ designate the electron and muon flavors respectively. The nonzero mixing angle ($\sin^2\theta\approx0.3$~\cite{PDG2020}) is responsible for the customary, neutrino flavor oscillations $\nu_{e(\mu)}^{L}\leftrightarrow\nu_{\mu(e)}^{L}$. 
In the presence of a magnetic field, the interaction of neutrino magnetic moments of diagonal ($\mu_{11}$ and $\mu_{22}$) and transition ($\mu_{12}$) types with a magnetic field induces the neutrino spin $\nu_{e(\mu)}^{L}\leftrightarrow\nu_{e(\mu)}^{R}$ and spin-flavor $\nu_{e(\mu)}^{L}\leftrightarrow\nu_{\mu(e)}^{R}$ oscillations. The exact solution of the problem in the case of a constant magnetic field $\vec{B}$ can be found in our earlier work~\cite{kurashvili2017spin}. Here, we wish to take into account the presence of magnetic-field fluctuations and to develop a general approach based on the quantum resource theory for the treatment of neutrino flavor, spin and spin-flavor oscillations.     

Below, in Sec.~\ref{nu_evolution}, we formulate the Lindblad master equation~\cite{Lindblad1976} for neutrino evolution that accounts for the dissipative effect due to a stochastic magnetic-field component, which can be present in different neutrino propagation environments, for example, in such as the interstellar space~(see Refs.~\cite{PlanckCollaboration2016,PhysRevD.96.123528}). Then, in Sec.~\ref{sec:DE}, we outline basic properties of the analytical solution of the Lindblad master equation for the neutrino density matrix. The numerical results based on the obtained solution, which quantify the coherence effects in neutrino oscillations of various types, are presented and discussed in Sec.~\ref{results}. Throughout we use the units in which $\hbar=c=1$, unless otherwise specified.

\section{The neutrino evolution equation}
\label{nu_evolution}
%
%
%
The effective Hamiltonian of the problem is~\cite{kurashvili2017spin}
\begin{eqnarray}\label{Hamiltonian of the problem}
&& \hat{H}_{eff}=\hat{H}_{vac}
+\hat{H}_{B}.
\end{eqnarray}
Here $\hat{H}_{vac}$ is the vacuum part and the term $\hat{H}_{B}$ describes the neutrino interaction with a magnetic field.  
The vacuum Hamiltonian in the flavor basis~(\ref{flavor basis}) has the form 
\begin{equation}
\label{eq:HamiltonianFlavorRepresentation}
{\hat H}_{vac}=\omega_\nu
\begin{pmatrix}
-\cos 2 \theta & 0 & \sin 2 \theta & 0
\\
0 & -\cos 2 \theta & 0 & \sin 2 \theta
\\
\sin 2 \theta & 0 & \cos 2\theta & 0
\\
0 & \sin 2 \theta & 0 & \cos 2\theta
\end{pmatrix},
\end{equation}
with
\begin{equation}
\label{eq:DeltaM}
\omega_\nu= \frac{\Delta m^2}{4 E_\nu}, \qquad \Delta m^2 =m_2^2 - m_1^2,
\end{equation}
and $E_\nu$ being the neutrino energy. 

The Hamiltonian of the neutrino interaction with a magnetic field in the
flavor representation can be presented as~\cite{fabbricatore2016neutrino}
%
\begin{widetext}
\begin{equation}
\label{eq:H_EM}
\hat{H}_{B}=
\begin{pmatrix}
\displaystyle -\left(\frac{\mu}{\gamma}\right)_{ee} {B_\parallel} &&
\mu_{ee}B_{\perp} &&
\displaystyle -\left(\frac{\mu}{\gamma}\right)_{ e\mu}{B_\parallel} &&
\mu_{e\mu}B_\perp
\\
\mu_{ee}B_\perp &&
\displaystyle \left(\frac{\mu}{\gamma}\right)_{ee}{B_\parallel} &&
\mu_{e\mu} B_\perp &&
\displaystyle \left(\frac{\mu}{\gamma}\right)_{e\mu}{B_\parallel}
\\
\displaystyle -\left(\frac{\mu}{\gamma}\right)_{e\mu}{B_\parallel} &&
\mu_{e\mu}B_{\perp} &&
\displaystyle -\left(\frac{\mu}{\gamma}\right)_{\mu \mu}{B_\parallel}
&&
\mu_{\mu \mu} B_\perp
\\
\mu_{e\mu}B_{\perp} &&
\displaystyle \left(\frac{\mu}{\gamma}\right)_{e\mu}{B_\parallel} &&
\mu_{\mu \mu}B_\perp &&
\displaystyle \left(\frac{\mu}{\gamma}\right)_{\mu\mu}{B_\parallel}
\end{pmatrix},
\end{equation}
\end{widetext}
where $B_\parallel$ and $B_\perp$ are the parallel and transverse magnetic-field components with respect to the neutrino velocity, and the neutrino magnetic moments $\tilde{\mu}_{\ell\ell'}$ and $\mu_{\ell\ell'}$ ($\ell,\ell'=e,\mu$) are related to those in the mass representation $\mu_{jk}$ ($j,k=1,2$) as follows:
\begin{align}
\label{eq:MuPrime}
\mu_{ee}&=\mu_{11} \cos^2 \theta +\mu_{22} \sin^2 \theta
+\mu_{12} \sin 2\theta,
\nonumber
\\
\mu_{e\mu}&=\mu_{12}\cos 2\theta + \frac{1}{2}
\left( \mu_{22} - \mu_{11}\right)\sin 2\theta,
\nonumber
\\
\mu_{\mu\mu}&=\mu_{11} \sin^2 \theta
+\mu_{22} \cos^2 \theta-\mu_{12} \sin 2\theta,
\end{align}
and
\begin{align}
\label{eq:MuTilde}
\left(\frac{\mu}{\gamma}\right)_{ee} &=
\frac{\mu_{11}}{\gamma_{1}}\,\cos^2 \theta +
\frac{\mu_{22}}{\gamma_{2}}\,\sin^2 \theta +
\frac{\mu_{12}}{\gamma_{12}}\,\sin 2\theta,
\nonumber
\\
\left(\frac{\mu}{\gamma}\right)_{e\mu} &=
\frac{\mu_{12}}{\gamma_{12}}\,\cos 2\theta
+\frac{1}{2}\left(
\frac{\mu_{22}}{\gamma_{2}}-\frac{\mu_{11}}{\gamma_{1}}
\right)\sin 2\theta,
\nonumber\\
\left(\frac{\mu}{\gamma}\right)_{\mu\mu}&=
\frac{\mu_{11}}{\gamma_{1}}\,\sin^2 \theta
+\frac{\mu_{22}}{\gamma_{2}}\,\cos^2 \theta
-\frac{\mu_{12}}{\gamma_{12}}\,\sin 2\theta.
\end{align}
Here $\gamma_1$ and $\gamma_2$ are the Lorenz factors of the massive neutrinos, and
\begin{equation}
\label{eq:GammaDefinition}
\frac{1}{\gamma_{12}}=\frac{1}{2}\left(\frac{1}{\gamma_1}+\frac{1}{\gamma_2}\right).
\end{equation}

In addition to the usual deterministic part $\vec{B}$ that enters Eq.~(\ref{eq:H_EM}) we consider a stochastic magnetic field 
$\vec{h}$. The stochastic field is characterized by the correlation function~\cite{PhysRevB.55.3050} $\langle h_\alpha(t)h_\beta(0)\rangle=\frac{w^2}{2\mu_\nu^2}\delta_{\alpha\beta}\delta(t)$, where $\mu_\nu$ is a putative value of the neutrino magnetic moment and $w^2=k_BT$, with $T$ being the effective temperature. To describe the neutrino motion in a fluctuating magnetic field, we employ the Lindblad master equation, which is widely used in studies of neutrino quantum decoherence in different environments and under various  experimental conditions (see Ref.~\cite{Stankevich:2020prd} and references therein). The density matrix of the system thus obeys the following equation:
\begin{eqnarray}\label{master equation}
 \frac{d\hat{\varrho}}{dt}&=&-{i}\left[\hat H, \hat\rho\right] \nonumber\\
&{}& -\frac{w^{2}}{2}\,\tr\left(\hat{\varrho}\hat{V^2}+\hat{V^2}\hat{\varrho}-2\hat{V}\hat{\varrho}\hat{V}\right).
\end{eqnarray}
We analytically solve it in the eigenbasis $\vert\psi_{i=1,2,3,4}\rangle$ of the Hamiltonian 
$\hat{H}_{eff}$ (see Ref.~\cite{kurashvili2017spin} for details). 
The equation for the density matrix takes the form
\begin{align}
\frac{d\varrho_{nm}}{dt} = &-i (E_n - E_m)\varrho_{mn} - \frac{w^2}{2}
\sum_q \left(\varrho_{nq} V^2_{qm} + V^2_{nq} \varrho_{qn}\right)
\nonumber \\
& + w^2 \sum_{q,s} V_{nq} \rho_{qs} V_{sm},
\label{eq:lindblad}
\end{align}
where $E_{i=1,2,3,4}$ are the eigenenergies of the Hamiltonian $\hat{H}_{eff}$. The matrix $V$ has the following general form:
\begin{equation}
\label{eqn:Vmatrix}
V_{ik} = \bra{\psi_i} \hat{I}_1  \otimes \hat{v}_2 + \hat{I}_2 \otimes \hat{v}_1 \ket{\psi_k} ,
\end{equation}
where $\hat{v}$ is a $2 \times 2$ matrix, and the
subscripts 1, 2 denote the action of a matrix on the
space of the first and second massive neutrinos, respectively.
The matrix $\hat{v}$ can be expanded into the basis of $2\times 2$ unit matrix and
three Pauli matrices:
\begin{equation}
\hat{v} = v_0 \hat{I} + \vec v \cdot  \hat{\vec\sigma}.
\label{eq:vsigma}
\end{equation}

Let us present the density matrix as
\begin{equation}
\label{eq:dmatrix_R}
\hat\varrho =
\begin{pmatrix}
\hat\varrho^{(11)} & \hat\varrho^{(12)} \\
\hat\varrho^{(21)} & \hat\varrho^{(22)}
\end{pmatrix}.
\end{equation}
The quadrants $\hat\varrho^{(\alpha)}$ are $2\times 2$ minors of the full density
matrix and can be expanded in terms of the unit and Pauli matrices:
\begin{equation}
\hat\varrho^{(\alpha)} = r_0^{(\alpha)}\hat{I} + \vec r^{(\alpha)}\cdot\hat{\vec\sigma},
\label{eq:Rsigma}
\end{equation}
where the expansion coefficients are defined by
\begin{equation}
\label{eq:Rsigma_trace}
r_{i=1,2,3}^{(\alpha)} = \frac 1 2 \,\tr\{ \hat\varrho^{(\alpha)}\hat\sigma_{i=1,2,3} \}, \qquad
r_0^{(\alpha)} = \frac 1 2\, \tr\hat\varrho^{(\alpha)}.
\end{equation}
In Eq.~(\ref{eq:lindblad}), the dissipative term contains the following 
two matrix terms arising from the combinations of $\hat{v}$ and $\hat\varrho^{(\alpha)}$:
\begin{equation}
\label{eq:Vsqr}
\hat{L}^{(\alpha)}_1 =
\hat{v}^2 \hat\varrho^{(\alpha)} + \hat\varrho^{(\alpha)}\hat{v}^2
\end{equation}
and
\begin{equation}
\label{eq:VAV}
\hat{L}^{(\alpha)}_2=
 \hat{v} \hat\varrho^{(\alpha)}\hat{v}
\end{equation}
for the first and second sums, respectively.

We now transform Eqs.~(\ref{eq:Vsqr}) and~(\ref{eq:VAV}) using
Eqs.~(\ref{eq:vsigma}) and~(\ref{eq:Rsigma}):
\begin{align}
\label{eq:Vsqr_expand}
 \hat{L}^{(\alpha)}_1 =& 2 \left[ \left( v_0^2 + v^2 \right)r^{(\alpha)}_0 +
2 v_0 \vec v \cdot \vec r^{(\alpha)} \right] \hat{I} 
 \nonumber\\
& + 4 v_0 r^{(\alpha)}_0 \vec v \cdot \hat{\vec \sigma} +
2 \left( v_0^2 + v^2 \right) \vec r^{(\alpha)} \cdot \hat{\vec \sigma},
\end{align}
and
\begin{align}
\label{eq:VAV_expand}
 \hat{L}^{(\alpha)}_2 = &\left[ \left( v_0^2 + v^2 \right)r^{(\alpha)}_0
+ 2 v_0 \vec v \cdot \vec r^{(\alpha)} \right] \hat{I} 
 \nonumber\\
& +2 \left[ v_0 r^{(\alpha)}_0 + \vec v \cdot \vec r^{(\alpha)}\right]
\vec v \cdot \hat{\vec \sigma}
 \nonumber\\
& + \left( v_0^2 - v^2 \right) \vec r^{(\alpha)}\cdot \hat{\vec \sigma}.
\end{align}

Summing up Eqs.~(\ref{eq:Vsqr_expand}) and~(\ref{eq:VAV_expand})
with the same weights as in Eq.~(\ref{eq:lindblad}),
one gets the full dissipative term:
\begin{align}
\label{eq:dissipative}
 \hat{L}^{(\alpha)} &=
 - \frac {w^2} 2 \hat{L}^{(\alpha)}_1 + w^2 \hat{L}^{(\alpha)}_2
\nonumber\\
& = 2 w^2 \left[ (\vec v \cdot \vec r^{(\alpha)})
\vec v \cdot \hat{\vec \sigma} -
v^2  \vec r^{(\alpha)} \cdot \hat{\vec \sigma}\right].
\end{align}
We also decompose Eq.~(\ref{eq:dissipative}) in the
basis of $2 \times 2$ matrices:
\begin{equation}
      \label{eq:Ldecompose}
      \hat{L}^{(\alpha)} =  \Lambda^{(\alpha)}_0 \hat{I} +
      \vec \Lambda^{(\alpha)} \cdot \hat{\vec \sigma},
\end{equation}
where
\begin{align}
\label{eq:l0}
\Lambda^{(\alpha)}_0 & = 0, \\
\label{eq:l1}
\Lambda^{(\alpha)}_i & = 2 w^2
\left[(\vec v \cdot \vec r^{(\alpha)}) v_i - v^2 r_i^{(\alpha)} \right].
\end{align}

Using Eqs.~(\ref{eq:Rsigma}), (\ref{eq:Vsqr_expand}), (\ref{eq:VAV_expand}), (\ref{eq:dissipative}), and~(\ref{eq:lindblad}),
one gets the system of equations for the elements of the minor $\hat\varrho^{(11)}$:
\begin{align}
\label{eq:diff_rho11}
\frac{d}{dt}\varrho_{11} (t) & =
\frac d {dt} [r^{(11)}_0(t) + r^{(11)}_3(t)] = \Lambda^{(11)}_3(t),
\\
\label{eq:diff_rho22}
\frac{d}{dt} \varrho_{22}(t)& =
\frac d {dt} [r^{(11)}_0(t) - r^{(11)}_3(t)]= -\Lambda^{(11)}_3(t),
\\
\label{eq:diff_rho12}
\frac d {dt} \varrho_{12}(t) & =
\frac d {dt} r^{(11)}_-(t) = - i\omega_{12} r^{(11)}_-(t) + \Lambda^{(11)}_-(t),
\\
\label{eq:diff_rho21}
\frac d {dt} \varrho_{21}(t) & =
\frac d {dt} r^{(11)}_+ (t)= - i \omega_{21} r^{(11)}_+(t) 
+ \Lambda^{(11)}_+(t),
\end{align}
where $r_\pm = r_1 \pm i r_2$, $\Lambda_\pm = \Lambda_1 \pm \Lambda_2$
and $\omega_{12} = E_1 - E_2 = -\omega_{21}$.
Note that the sum of the diagonal matrix elements
$\varrho_{11}(t) +\varrho_{22}(t) = 2 r^{(11)}_0(t)$
is time-independent.

The set of equations for another ``diagonal'' minor, 
$\hat\varrho^{(22)}$, is obtained from
Eqs.~(\ref{eq:diff_rho11})-(\ref{eq:diff_rho21})
by changing $r^{(11)}_{i=0,1,2,3}$ to the corresponding $r^{(22)}_{i=0,1,2,3}$,
and $\omega_{12}$ to $\omega_{34}$. Similarly, $\varrho_{33}(t) + \varrho_{44}(t) = 2 r^{(22)}_0(t)$
is time-independent, as well as the complete
trace of the density matrix $\tr\hat{\varrho}(t)=2 [r^{(11)}_0(t)+r^{(22)}_0(t)]=1$.

The system of equations for the minor $\hat\varrho^{(12)}$ is
\begin{align}
\label{eq:diff_rho13}
\frac d {dt}\varrho_{13}(t) & = 
\frac d {dt} [r^{(12)}_0(t) + r^{(12)}_3(t)]
\nonumber\\
& =- i \omega_{13}[r^{(12)}_0(t) + r^{(12)}_3(t)]
+ \Lambda^{(12)}_3(t),
\\
\label{eq:diff_rho24}
\frac d {dt} \varrho_{24}(t) & = \frac d {dt} [r^{(12)}_0(t) - r^{(12)}_3(t)]
 \nonumber\\
& =- i \omega_{24}[r^{(12)}_0(t) - r^{(12)}_3(t)]  - \Lambda^{(12)}_3(t),
\\
\label{eq:diff_rho14}
\frac d {dt} \varrho_{14}(t) & =
\frac d {dt} r^{(12)}_-(t) =
- i \omega_{14} r^{(12)}_-(t)  + \Lambda^{(12)}_-(t),
\\
\label{eq:diff_rho23}
\frac d {dt} \varrho_{23}(t) & =
\frac d {dt} r^{(12)}_+(t) =
- i\omega_{23} r^{(12)}_+ (t) + \Lambda^{(2)}_+(t).
\end{align}
Finally, the equations for the matrix elements of $\hat\varrho^{(21)}$ are obtained from
Eqs.~(\ref{eq:diff_rho13})-(\ref{eq:diff_rho23}) by means of Hermitian
conjugation.

\section{Solution of the master equation}
\label{sec:DE}
%
%
For illustrative purposes we assume that
\begin{equation}
\mu_{11}=\mu_{22}=\mu_{12} = \mu_\nu.
\label{eqn:momentum_equality}
\end{equation}
Since neutrinos are ultrarelativistic particles, the terms involving Lorentz factors in Eq.~(\ref{eq:H_EM}) can be safely neglected. In what follows, we set $B_\perp=B$. 
%
The resulting effective Hamiltonian~(\ref{Hamiltonian of the problem}) has the following characteristic equation:
\begin{equation}
E^4 - 2(\omega_\nu^2 +2 \mu_\nu^2 B^2) E^2 + \omega_\nu^4 =0.
\label{eq:characteristic_hamiltonian}
\end{equation}
Its roots are given by
\begin{equation}
E_{1,2,3,4} = \mp  \left( 
\sqrt{\omega_\nu^2 + {\mu_\nu^2 B^2 }} 
\pm {\mu_\nu B}\right),
\label{eq:characteristic_hamiltonian_solution}
\end{equation}
where the eigenenergies $E_{1,2}$ ($E_{3,4}$) correspond to the minus (plus) sign in front of the brackets. Let us define  
the energy splitting $\omega_B$ due to the presence of a magnetic field:
\begin{equation}
\omega_{B} =  E_1 - E_2 = E_3-E_4 = 2\mu_\nu B.
\label{eqn:omegaB}
\end{equation}
This energy value determines the characteristic frequency scale of neutrino spin oscillations, while that for flavor oscillations is represented by the modified frequency 
\begin{equation}
\omega_{N} = \frac {\omega_{31} + \omega_{42}} 2 =
\sqrt{\omega_\nu^2 + \mu_\nu^2 B^2 }.
\label{eq:omegaN}
\end{equation}

Consider now the system of equations for the minor $\hat\varrho^{(11)}$. One can rewrite Eqs.~(\ref{eq:diff_rho11})-(\ref{eq:diff_rho21}) as
\begin{align}
\label{eq:diff_r0}
\frac d {d\tau} r_0^{(11)}(\tau)  = &0, \\
\label{eq:diff_rplus}
\frac d {d\tau} r_+^{(11)} (\tau) =&
\left(\frac{v_+ v_-} 2 - v_3^2 - i \homega \right) r_+^{(11)}(\tau)
+ \frac{v_+^2} 2 r_-^{(11)}(\tau)
\nonumber\\
&  +v_+ v_3 r_3^{(11)}(\tau), \\
\label{eq:diff_rminus}
\frac d {d\tau} r_-^{(11)}(\tau)  =&
\frac{v_-^2} 2 r_+^{(11)}(\tau)
+ \left( - \frac{v_+ v_-} 2 -  v_3^2 + i \homega \right) r_-^{(11)}(\tau)
\nonumber\\
&    +v_- v_3 r_3^{(11)}(\tau), \\
\label{eq:diff_r3}
\frac d {d\tau} r_3 (\tau) =&
\frac{v_-v_3}{2} {r_+^{(11)}(\tau)} + \frac{v_+v_3}{2} {r_-^{(11)}(\tau)}\nonumber\\
& - v_+ v_- r_3^{(11)}(\tau),
\end{align}
where $v_\pm=v_1\pm iv_2$ and we introduced the reduced time variable
$\tau =  2 w^2  t$ 
and the reduced frequency $\homega = \omega_{12}/2w^2=\mu_\nu B/w^2=\homega_B$.

The relevant components of the matrix $\hat{v}$ are expressed by a three-dimensional vector $\vec v$
[as can be seen above, the component $v_0$ is no longer relevant because it does not appear in
the final expression for the dissipative term in Eq.~(\ref{eq:dissipative})], which can be parametrized as
\begin{equation}
\label{eq:VAlphaBeta}
\vec v = (  v\cos \varphi \sin \beta,  v\sin \varphi \cos \varphi ,  v\cos \beta),
\end{equation}
where $\varphi$ and $\beta$ are some angle parameters. Since the norm $v$ appears in the combination with $w^2$, it
can be included in the definition of the latter parameter and set to $v=1$. We also assume the matrix $\hat v$ to be real, setting $\varphi=0$, and hence
\begin{equation}
\label{eq:VBeta}
\vec v = (\sin \beta, 0 , \cos \beta).
\end{equation}

Using $r_{1,2}^{(11)}$ instead of $r_\pm^{(11)}$ in Eqs.~(\ref{eq:diff_r0})-(\ref{eq:diff_r3}), we get
%
%
\begin{align}
\label{eqn:diff_r0_alt_coherence}
\frac d {d\tau} r^{(11)}_0 & = 0,
\\
\label{eqn:diff_r1_alt_coherence}
\frac d {d\tau} r^{(11)}_1 & = -r^{(11)}_1\cos^2 \beta  + 
r^{(11)}_2 \bar \omega_{B}  + 
r^{(11)}_3 \sin \beta \cos \beta ,
\\
\label{eqn:diff_r2_alt_coherence}
\frac d {d\tau} r^{(11)}_2 & = - r^{(11)}_1 \bar \omega_{B}
- r^{(11)}_2 \cos^2 \beta ,
\\
\label{eqn:diff_r3_alt_coherence}
\frac d {d\tau} r^{(11)}_3 & =  r^{(11)}_1 \sin \beta \cos \beta -
r^{(11)}_3 \sin^2 \beta.
\end{align}
The system of equations for the variables $r^{(22)}_{i=0,1,2,3}(t)$ is obtained in a similar manner.

In the case of the minor $\hat\varrho^{(12)}$ we deduce from Eqs.~(\ref{eq:diff_rho13})-(\ref{eq:diff_rho23}):
\begin{align}
\label{eqn:diff_r0_2_text}
\frac d {d\tau} r^{(12)}_ 0 =  &- i  \homega_+ r^{(12)}_ 0 - 
i \homega_- r^{(12)}_3,
\\
\label{eqn:diff_rp_2_text}
\frac d {d\tau} r^{(12)}_1 = &
( - \cos^2 \beta - i \homega_+) r^{(12)}_1  +
\homega_0 r^{(12)}_2 +
 \nonumber\\
& + r^{(12)}_3 \sin \beta \cos \beta,
\\
\label{eqn:diff_rm_2_text}
\frac d {d\tau} r^{(12)}_2 = &
- \homega_0  r^{(12)}_1 + ( -\cos^2 \beta - i \homega_+)r^{(12)}_2,
\\
\label{eqn:diff_r3_2_text}
\frac d {d\tau} r^{(12)}_3 = &
-i \homega_- r^{(12)}_0 + r^{(12)}_1 \sin \beta \cos \beta 
 \nonumber\\
& - ( i\homega_+  + \sin^2 \beta)r^{(12)}_3.
\end{align}
The reduced frequencies are given by 
\begin{align}
\label{eqn:freq0_text}
\homega_0 & = \frac{\homega_{12} + \homega_{34} }{2} =
\homega_{B},
\\
\label{eqn:freqpm_text}
\homega_\pm & = \frac{\homega_{31} \pm \homega_{42} }{2}
= \frac{\homega_{N} \pm \homega_{N}} 2,
\end{align}
where $\homega_{ij}=\omega_{ij}/2w^2$ and $\homega_{N}=\omega_{N}/2w^2$. Utilizing the substitution
\begin{equation}
r^{(12)}_i (\tau) = e^{-i\homega_{N} \tau} R^{(12)}_i (\tau)
\label{eqn:substitution_R2}
\end{equation}
in Eqs.~(\ref{eqn:diff_r0_2_text})-(\ref{eqn:diff_r3_2_text}) and taking into account
Eqs.~(\ref{eqn:freq0_text}) and~(\ref{eqn:freqpm_text}), we find 
that functions $R^{(12)}_i(\tau)$ obey exactly the same system of equations as that given by 
Eqs.~(\ref{eqn:diff_r0_alt_coherence})-(\ref{eqn:diff_r3_alt_coherence}).


In Eqs.~(\ref{eqn:diff_r0_alt_coherence})-(\ref{eqn:diff_r3_alt_coherence}) the first equation is trivial and therefore the
solution of the system requires diagonalization of the $3\times3$ matrix
\begin{equation}
\label{eq:MR1}
\mathcal{M}^{(1)}_3 =
\begin{pmatrix}
- \cos^2\beta && \homega_B && \sin\beta \cos\beta
\\
-\homega_B && - \cos^2\beta  && 0
\\
\sin\beta\cos\beta  && 0 && -\sin^2\beta
\end{pmatrix}.
\end{equation}
The general solution to the system is a sum of exponents:
\begin{equation}
\label{eq:solution_R1}
r^{(11)}_i (\tau) = \sum_{k=1}^{3} C_{ik} e^{\zeta_k \tau},
\end{equation}
where $\zeta_i$ are the eigenvalues of the matrix~(\ref{eq:MR1}).
The integration constants are given by the following
expressions:
\begin{align}
\label{eq:R1_c1}
C_{i1} & = \frac {B_{0i} \zeta_2 \zeta_3 - B_{1i} (\zeta_2 + \zeta_3)
+ B_{2i}}
{(\zeta_1 - \zeta_2)(\zeta_1 - \zeta_3)},\\
\label{eq:R1_c2}
C_{i2} & = \frac{B_{0i} \zeta_1 \zeta_3 - B_{1i} (\zeta_1 +\zeta_3)
+B_{2i}}
{(\zeta_2 - \zeta_1)(\zeta_2 - \zeta_3)},\\
\label{eq:R1_c3}
C_{i3} & = \frac{B_{0i} \zeta_1 \zeta_2 - B_{1i} (\zeta_1 + \zeta_2)
+B_{2i}}
{(\zeta_3 - \zeta_1)(\zeta_3 -\zeta_2)},
\end{align}
where
\begin{align}
\label{eq:R1betas}
B_{0i} & = r^{(11)}_i(0), 
\\
B_{1i} & = \sum_k \mathcal{M}^{(1)}_{3\,ik}r^{(11)}_{k}(0), 
\\
B_{2i} & = \sum_{kl} \mathcal{M}^{(1)}_{3\,ik}
\mathcal{M}^{(1)}_{3\,kl}r^{(11)}_{l}(0).
\end{align}

The characteristic equation for the system of 
Eqs.~(\ref{eqn:diff_r0_alt_coherence})-(\ref{eqn:diff_r3_alt_coherence})
is
\begin{equation}
\label{eqn:characteristic_MR1_coherence}
\zeta^3 + (2-\sin^2 \beta)\zeta^2 +(\homega_B^2 + \cos^2 \beta) \zeta
+\homega_B^2 \sin^2 \beta =0.
\end{equation}
It has three roots: the one ($\zeta_3$) is always real and the other two are complex conjugate to each other ($\zeta_1=\zeta_2^*$).

Let us briefly outline properties of the solution in the two limiting cases $\beta=0$ and $\pi/2$. 
In the $\beta=0$ case, the matrix $\hat{v}$ has a diagonal form,
$\hat{v}=\hat{\sigma}_z$. The roots of the characteristic equation~(\ref{eqn:characteristic_MR1_coherence}) are given by $\zeta_{1,2} = -1\pm i \homega_B$ and $\zeta_3=0$.
From Eqs.~(\ref{eqn:diff_r0_alt_coherence})-(\ref{eqn:diff_r3_alt_coherence}) it follows that 
the ``longitudinal'' component $r^{(11)}_3$ is time-independent and the ``transverse'' components $r^{(11)}_1$ and $r^{(11)}_2$ decay as $\propto e^{-\tau}$, oscillating with frequency $\homega_B$.
If one sets them zero in the initial moment of time $\tau=0$ the density matrix remains constant and diagonal 
for all times $\tau > 0$. In the $\beta=\pi/2$ case, the real root of the characteristic
equation~(\ref{eqn:characteristic_MR1_coherence}) is $\zeta_3 = -1$, 
and the other two are purely imaginary: $\zeta_{1,2} =  \pm i\homega_B$. The ``longitudinal'' component decays, $r^{(11)}_3(\tau)\propto e^{-\tau}$, and the ``transverse'' components $r^{(11)}_1$ and $r^{(11)}_2$ oscillate with frequency $\homega_B$. 

\begin{figure}
\centering
\includegraphics[width=0.45\textwidth]{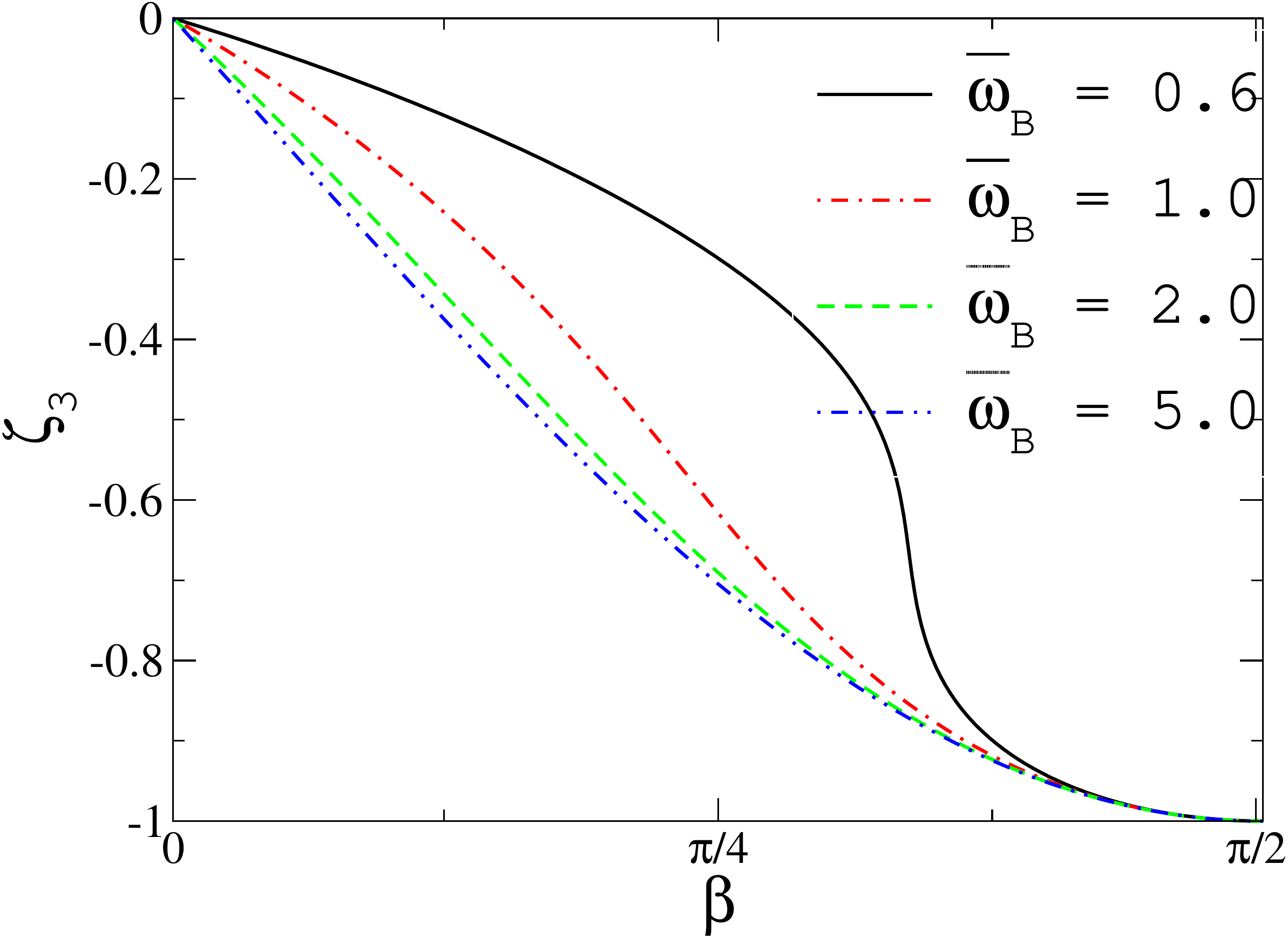}
\caption{The real root of the cubic equation~(\ref{eqn:characteristic_MR1_coherence}) as a function of angular parameter $\beta$ for different values of the reduced frequency $\homega_B=\omega_B/2w^2$. 
}
\label{fig:Re_nu3_theta}
\end{figure}

\begin{figure}
\centering
\includegraphics[width=0.45\textwidth]{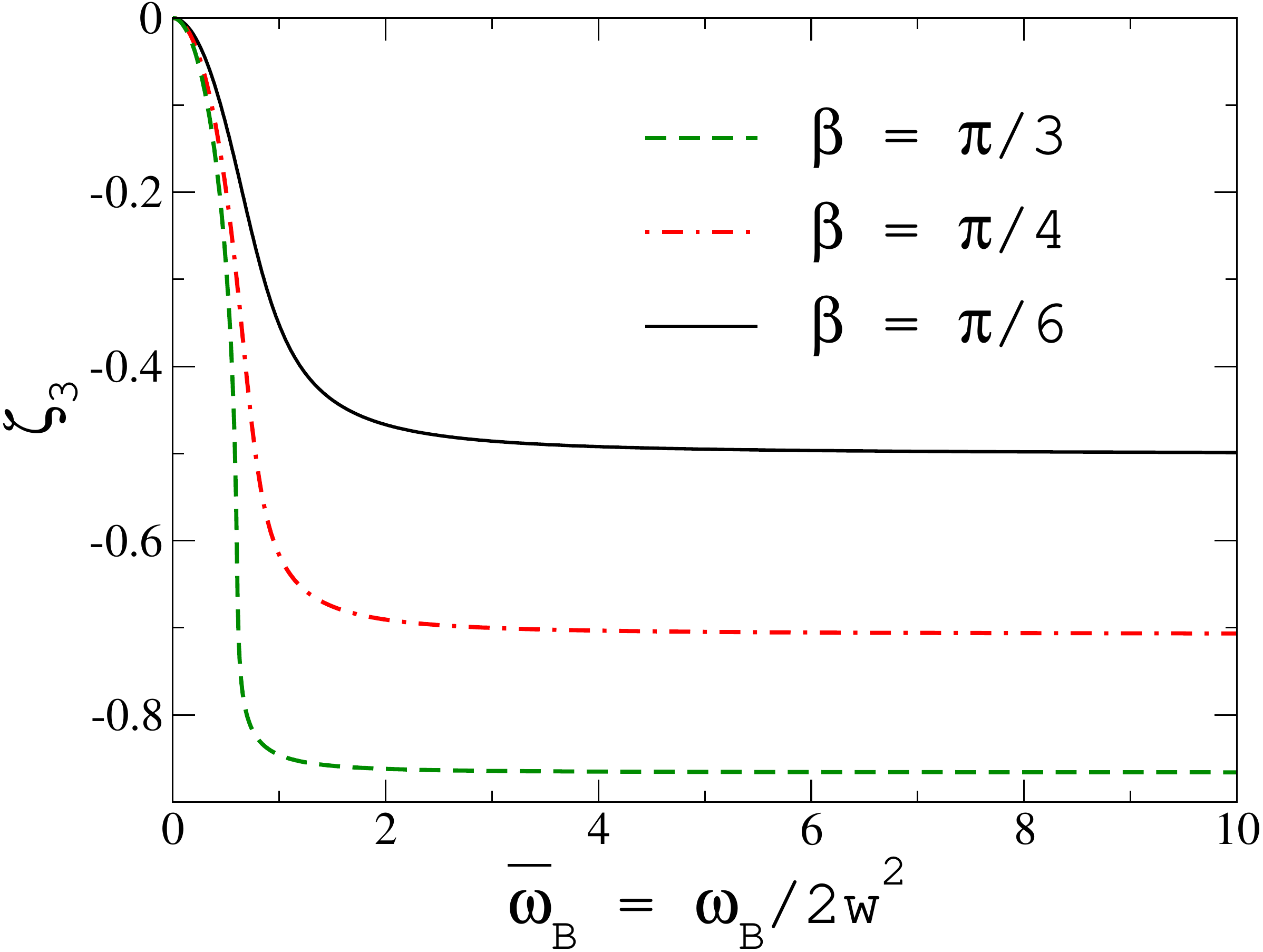}
\caption{The real root of the cubic equation~(\ref{eqn:characteristic_MR1_coherence}) as a function of the reduced frequency $\homega_B$ for particular values of the angle $\beta$. Each $\beta$ is characterized by the asymptotic value of
$\zeta_3$.
}
\label{fig:Re_nu3}
\end{figure}

\begin{figure}
\centering
\includegraphics[width=0.45\textwidth]{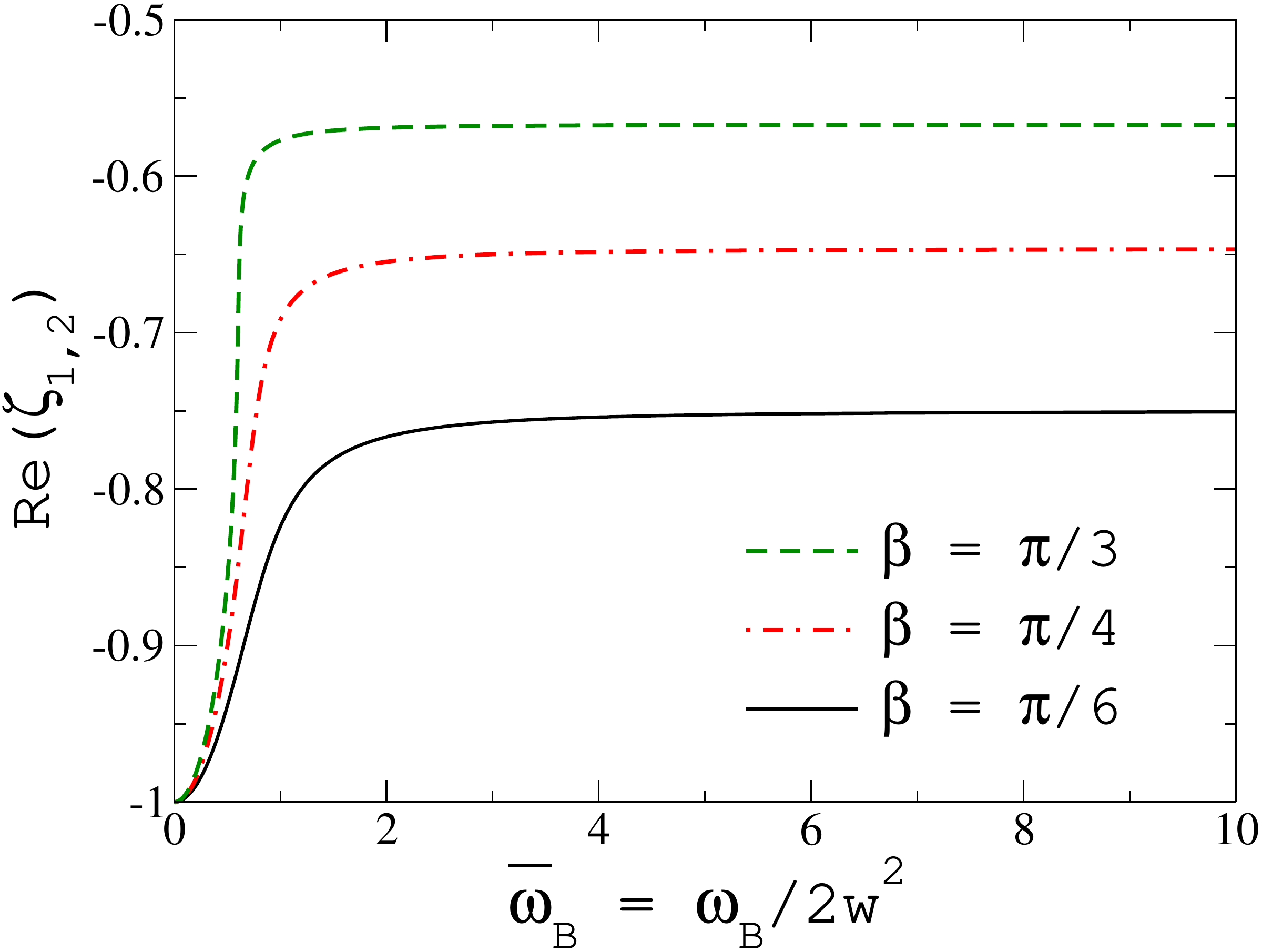}
\caption{The real part of the two complex
roots of the cubic equation~(\ref{eqn:characteristic_MR1_coherence}) as a function of the reduced frequency $\homega_B$ for particular values of the angle $\beta$. 
}
\label{fig:Re_nu12}
\end{figure}

When $0 < \beta < \pi/2$, the solution is more involved. Figure~\ref{fig:Re_nu3_theta} shows the dependence of the real root $\zeta_3$
on the angle $\beta$. One can see that $\zeta_3$ monotonically decreases from 0 to ${}-1$ with increasing $\beta$ from 0 to $\pi/2$, and the indicated decrease is faster for larger $\homega_B$ values. This means that the dissipation effect associated with the $\zeta_3$ term in Eq.~(\ref{eq:solution_R1}) is stronger for larger values of $\beta$ and $\homega_B$. From Fig.~\ref{fig:Re_nu3} it can be seen that at fixed value of $\beta$ the real root $\zeta_3$ drops from 0 to some asymptotic value with increasing $\homega_B$. In agreement with Fig.~\ref{fig:Re_nu3_theta}, the modulus of the asymptotic value is larger for larger $\beta$. This observation is  opposite to the behavior of the real part of two complex roots $\zeta_{1, 2}$ shown in Fig.~\ref{fig:Re_nu12}. As follows from the results presented  in Fig.~\ref{fig:Re_nu12}, the dissipation effect associated with the $\zeta_{1,2}$ terms in Eq.~(\ref{eq:solution_R1}) appears to be weaker for larger values of $\beta$ and $\homega_B$, in contrast to the $\zeta_3$ case.

%

\begin{figure}
\centering
\includegraphics[width=0.45\textwidth]{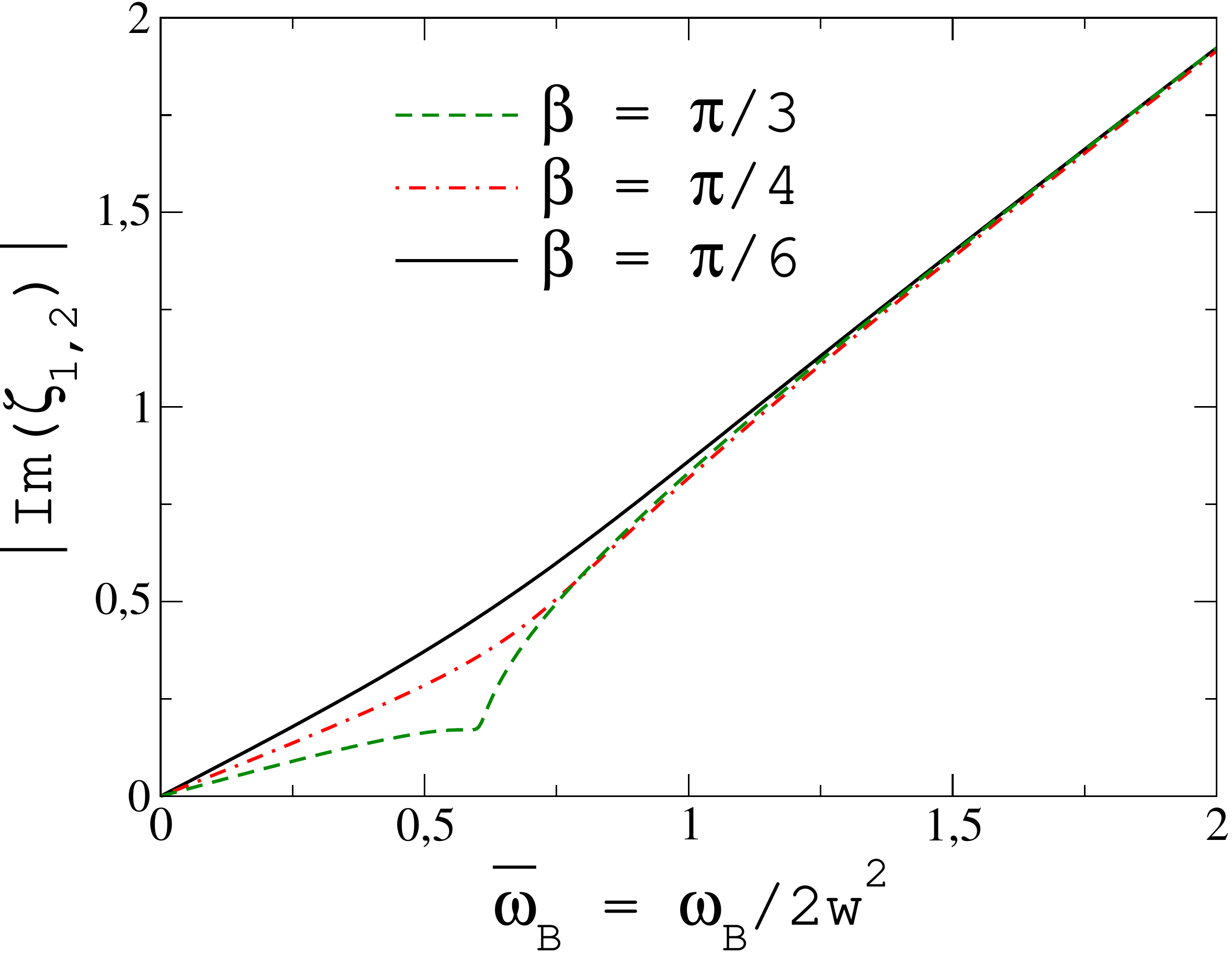}
\caption{The modulus of the imaginary parts of two complex roots of the cubic equation~(\ref{eqn:characteristic_MR1_coherence}) as a function of the reduced frequency $\homega_{B}$ for selected $\beta$ values. The curves corresponding to different values of the angle $\beta$ merge when $\homega_B$ is large.}
\label{fig:im12}
\end{figure}

%
The behavior of the imaginary parts of the complex roots $\zeta_{1,2}$ as functions of $\homega_B$ is
shown in Fig.~\ref{fig:im12}. It can be seen that the modulus of the imaginary parts grows with increasing $\homega_B$ and the curves corresponding to distinct $\beta$ values merge to nearly a linear function at large values of $\homega_B$. The proportionality coefficient asymptotically approaches unity. Note that, as discussed above, the dependence is exactly linear for $\beta=0,\pi/2$.

\begin{figure}
\centering
\includegraphics[width=0.45\textwidth]{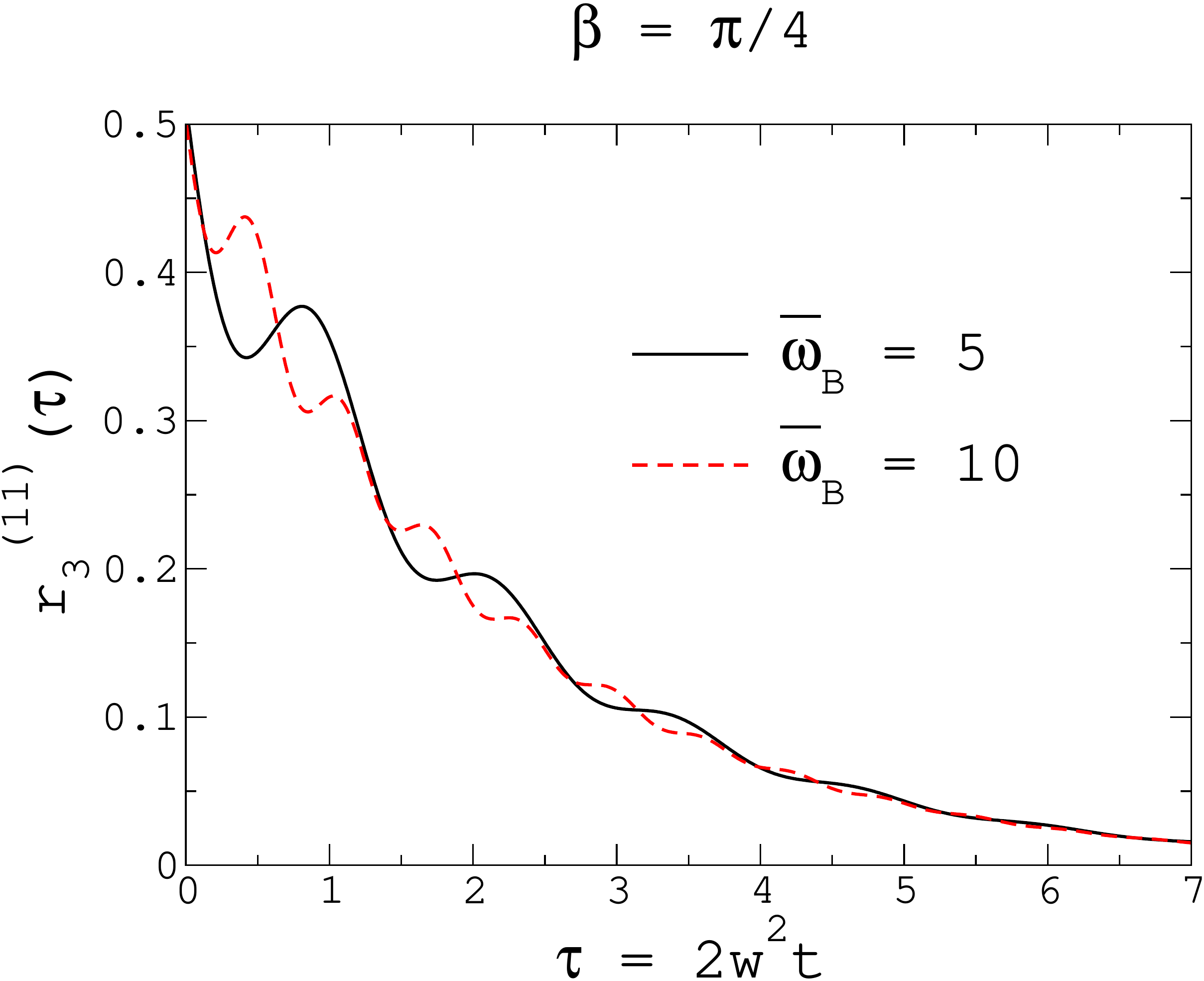}
\caption{The function $r^{(11)}_3(\tau)$ for two different values of the reduced frequency $\bar\omega$ and a fixed value $\beta=\pi /4$. The initial conditions are $r^{(11)}_0(0)=1/2$, $r^{(11)}_\pm(0) = 0$, $r^{(11)}_3(0)= 1/2$,
meaning that $\varrho^{(11)}_{11}(0) = \varrho_{11}=1$. 
The results are in line with the asymptotic behaviors of $\zeta_3$ and ${\rm Re}(\zeta_{1,2})$ in
Figs.~\ref{fig:Re_nu3} and~\ref{fig:Re_nu12} respectively.} 
\label{fig:Fig6}
\end{figure}

\begin{figure}[ht]
\centering
\includegraphics[width=0.45\textwidth]{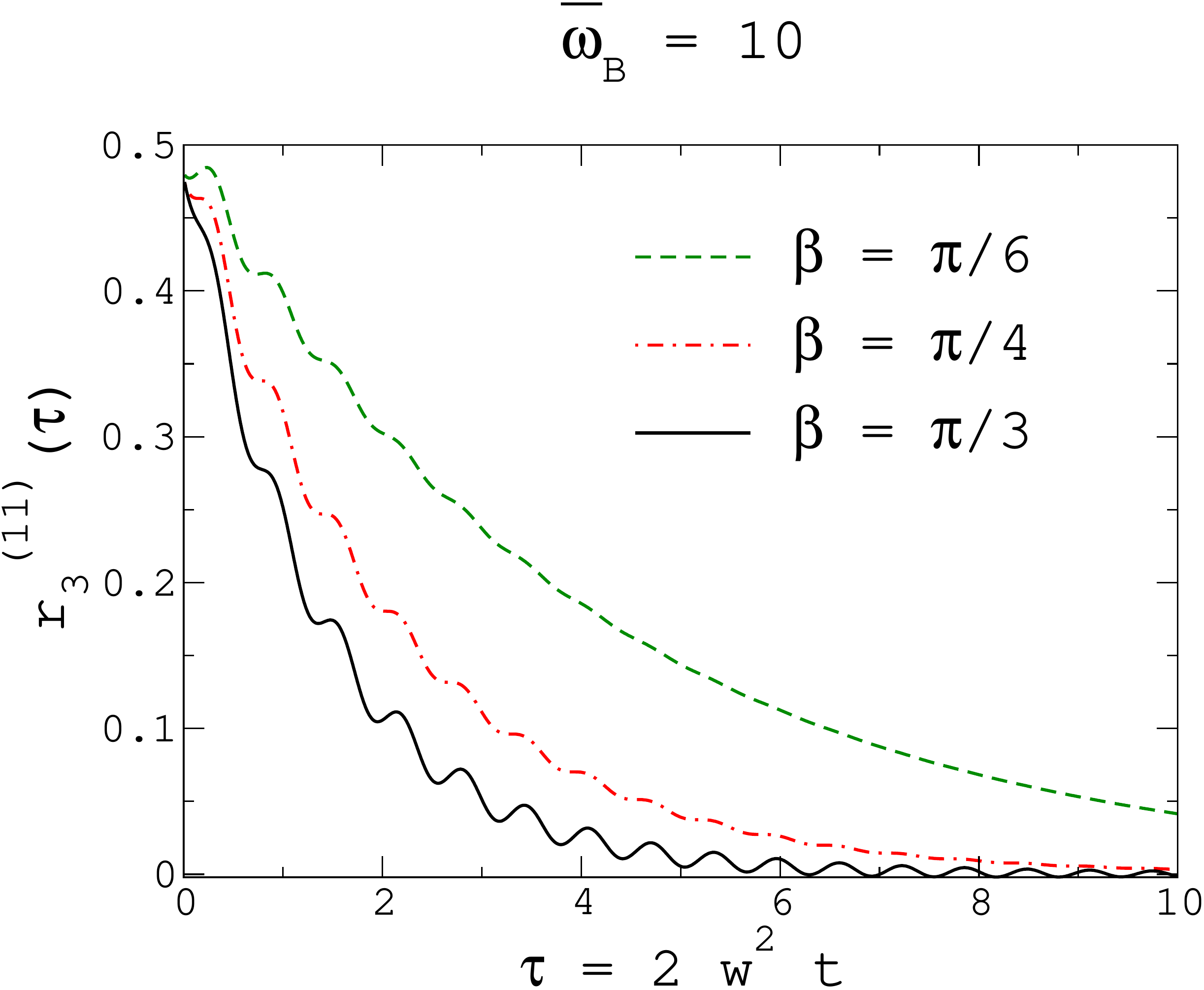}
\caption{Time dependence of the density-matrix component $r^{(11)}_3$ for the reduced frequency $\bar\omega=\mu_\nu B/2w^2 =10$ and three different values of the angular parameter $\beta$. 
The initial conditions are the same as in Fig.~\ref{fig:Fig6}.
}
\label{fig:Three_angles}
\end{figure}

The Lindblad equation parameter $w^2$ characterizes the strength of the dissipation effects and is usually equal to some fraction of the energy of particle interaction with a magnetic field. In our case, this fraction is determined by $w^2/\mu_\nu B =1/ \homega_B$.
Figure~\ref{fig:Fig6} shows the time evolution of the density-matrix component $r^{(11)}_3$ for an intermediate value of $\beta$, namely $\beta=\pi/4$. The result of the calculation demonstrates that the functions $r^{(11)}_3(\tau)$ for two different frequencies, $\homega_B=5$ and 10, decay at the same rate. This feature agrees with the results presented in Figs.~\ref{fig:Re_nu3} and~\ref{fig:Re_nu12}. The behaviors of time evolution of $r^{(11)}_3$ at a fixed $\homega_B$ value for different values of $\beta$ is shown in Fig.~\ref{fig:Three_angles}. It is clearly seen that the dissipation effect is $\beta$-dependent and is stronger for larger $\beta$. This observation is in line with results in Fig.~\ref{fig:Re_nu3}, but not with those in Fig.~\ref{fig:Re_nu12}. It points out that the decay of the $r^{(11)}_3(\tau)$ function is dominated by the $\zeta_3$ term rather than the $\zeta_{1,2}$ terms in Eq.~(\ref{eq:solution_R1}).

\section{Neutrino oscillations' probability, coherence and mixedness}
\label{results}
%
%
%
Suppose that at the initial moment of time $t=0$ the neutrino is in the active, left-handed electron-flavor state $\ket{\nu_e^L}$. The neutrino oscillations are typically characterized by the probabilities of the corresponding transitions $P_{\nu_e^L\to\nu_\mu^L}(t)$ (flavor),  $P_{\nu_e^L\to\nu_e^R}(t)$ (spin), and $P_{\nu_e^L\to\nu_\mu^R}(t)$ (spin-flavor), and by the survival probability $P_{\nu_e^L\to\nu_e^L}(t)$. These probabilities can be expressed in terms of the neutrino density matrix $\hat\varrho(t)$ as follows:
\begin{align}
\label{probability:survival}
P_{\nu_e^L\to\nu_e^L}(t)&=\tr\left\{\hat\varrho(t)\ket{\nu_e^L}\bra{\nu_e^L}\right\}=\rho_{d22}(t),\\
\label{probability:flavor}
P_{\nu_e^L\to\nu_\mu^L}(t)&=\tr\left\{\hat\varrho(t)\ket{\nu_\mu^L}\bra{\nu_\mu^L}\right\}=\rho_{d44}(t),\\
\label{probability:spin}
P_{\nu_e^L\to\nu_e^R}(t)&=\tr\left\{\hat\varrho(t)\ket{\nu_e^R}\bra{\nu_e^R}\right\}=\rho_{d11}(t),\\
\label{probability:spin-flavor}
P_{\nu_e^L\to\nu_\mu^R}(t)&=\tr\left\{\hat\varrho(t)\ket{\nu_\mu^R}\bra{\nu_\mu^R}\right\}=\rho_{d33}(t),
\end{align}
where $\rho_{dnn}$ are diagonal elements of the neutrino density matrix $\hat\varrho(t)$ in the flavor basis~(\ref{flavor basis}). The initial density matrix $\hat\varrho(0)$ contains only the left-handed electron neutrino, meaning that $\hat\varrho(0)=\rho_{d22}(0)\vert\nu_{e}^{L}\rangle\langle\nu_{e}^{L}\vert$, with $\rho_{d22}(0)=1$. 

\begin{figure}[ht]
\includegraphics[width=0.45\textwidth]{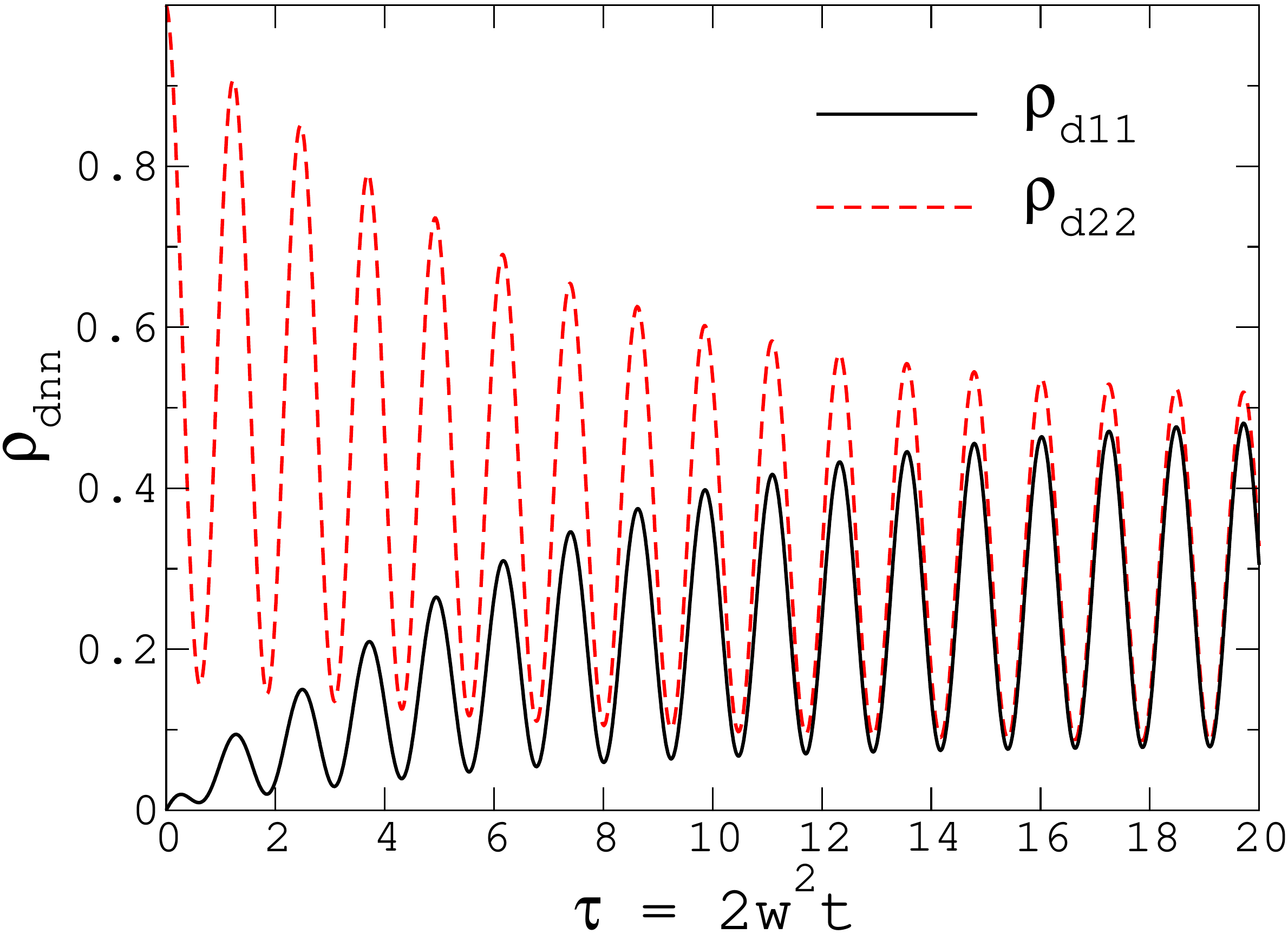}
\caption{The time dependence of $\rho_{d22}$ and $\rho_{d11}$ when $\beta=\pi/6$, $\homega_B=10$ and $\mu_\nu B /\omega_\nu =5$.  In the initial moment of time $\rho_{d11}(0)=1$ and all other elements of the density matrix $\hat\varrho(0)$ in the flavor representation are zero.}
\label{fig:diagonals12}
\end{figure}
\begin{figure}[ht]
\includegraphics[width=0.45\textwidth]{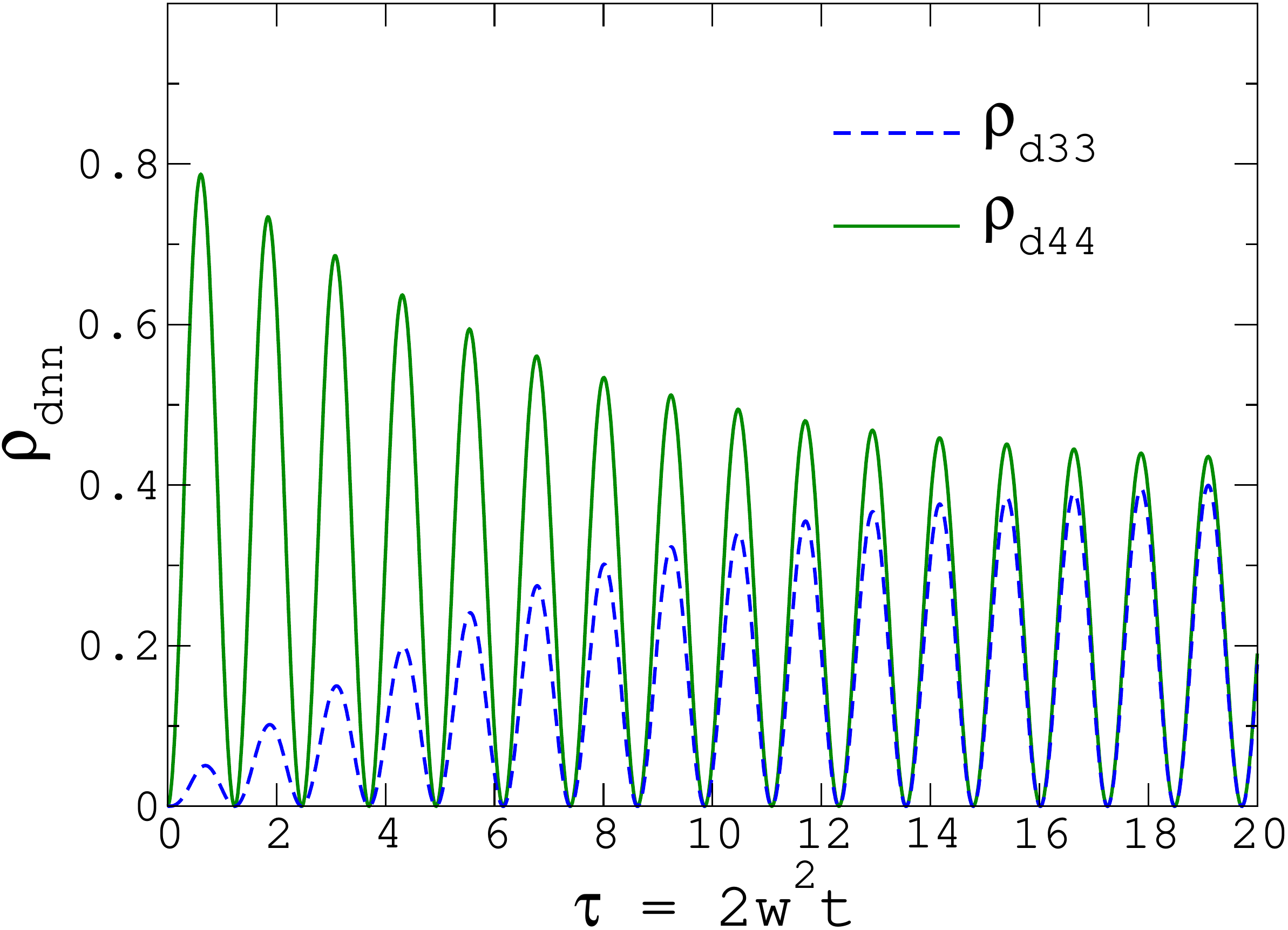}
\caption{The same as in Fig.~\ref{fig:diagonals12}, but for $\rho_{d44}$ and $\rho_{d33}$.}
\label{fig:diagonals34}
\end{figure}

The neutrino transition probabilities~(\ref{probability:survival})-(\ref{probability:spin-flavor}) are shown in Figs.~\ref{fig:diagonals12} and~\ref{fig:diagonals34}. As we can see, the spin-up and spin-down states of both flavors
thermalize, leading to an unpolarized steady state. On the other hand, we see a clear dominance of the electron neutrino in the final steady state: $\rho_{d11}(t)+\rho_{d22}(t)>\rho_{d33}(t)+\rho_{d44}(t)$. This dominance reflects the memory of the system about the initial neutrino flavor. In the case of the initial muon neutrino state (not shown), i.e., $\hat\varrho(0)=\rho_{44}(0)\vert\nu_{\mu}^{L}\rangle\langle\nu_{\mu}^{L}\vert$, with $\rho_{44}(0)=1$, we find the same memory effect of the initial netrino flavor. 


We now wish to quantify coherence of neutrino oscillations. For this purpose we employ the relative entropy as an entropic measure of such coherence~\cite{Sunil,Gerardo}:
\begin{equation}\label{coherence_measure}
\mathcal{C}\left(\hat{\varrho}(t)\vert\hat{\varrho}_d(t)\right)=\tr\left\{\hat{\varrho}(t)\ln\hat{\varrho}(t)-\hat{\varrho}(t)\ln\hat{\varrho}_d(t)\right\}.
\end{equation}
Here $\hat{\varrho}_d(t)$ is the diagonal part of $\hat{\varrho}(t)$ in the flavor basis~(\ref{flavor basis}), i.e., $\hat{\varrho}_d=\diag(\rho_{d11},\rho_{d22},\rho_{d33},\rho_{d44})$. Further, we use mixedness~\cite{Rastegin,Kumar}
\begin{equation}
\label{mixedness}
M(\hat{\varrho}(t))=\frac{d}{d-1}\left(1-\tr\hat{\varrho}^2\right), 
\end{equation}
where in our case $d=4$, and study the trade-off relation between coherence and mixedness:
\begin{equation}
\label{the trade off relation}
\frac{\mathcal{C}^{2}\left(\hat{\varrho}(t)\vert\hat{\varrho}_d(t)\right)}{(d-1)^{2}}+M(\hat{\varrho}(t))\leqslant 1.
\end{equation}

\begin{figure}[ht]
\includegraphics[width=0.45\textwidth]{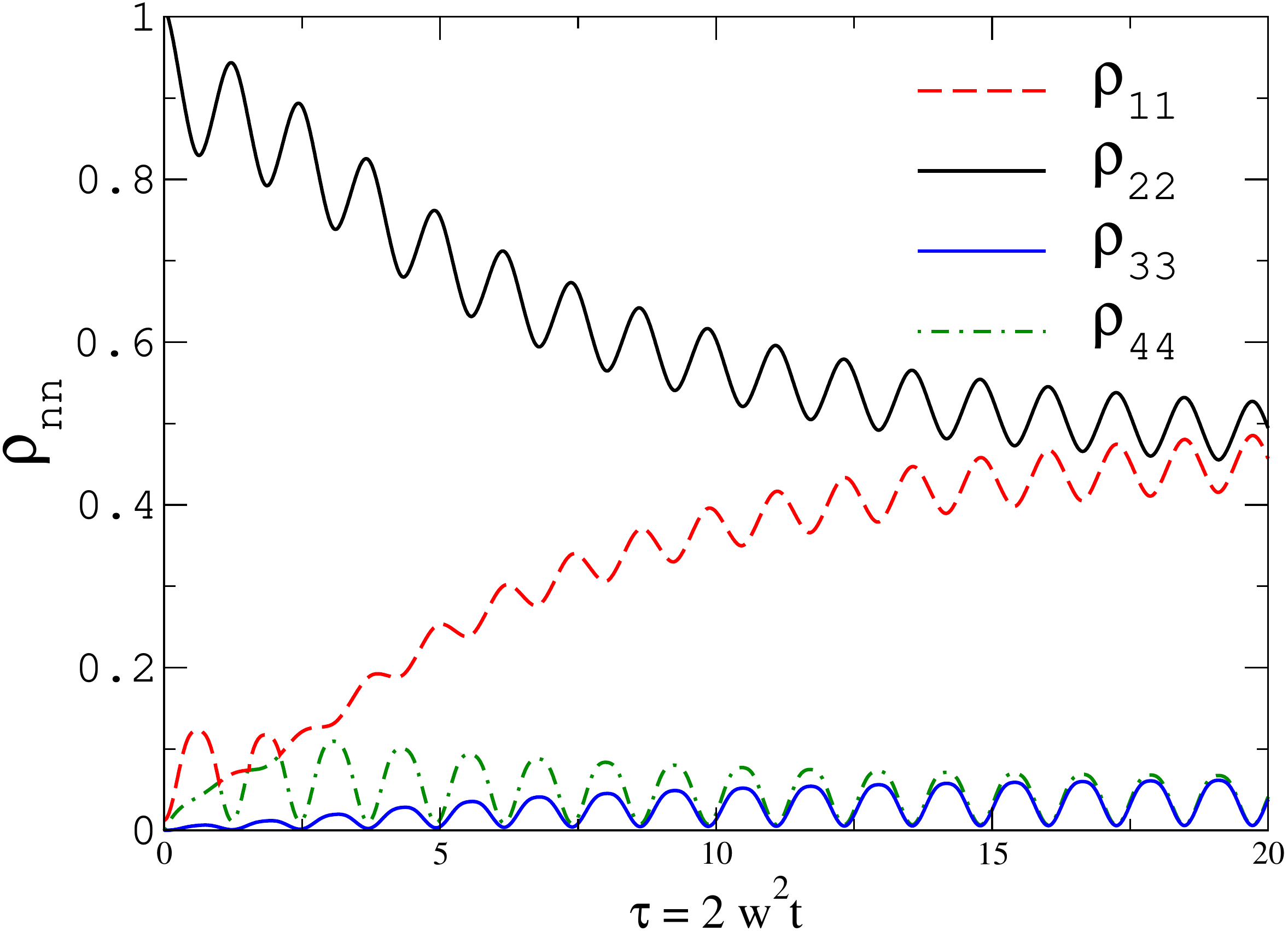}
\caption{The time dependence of eigenvalues $\rho_{nn}$ of the density matrix $\hat \varrho$
when $\beta = \pi/6$, $\homega_B=\mu_\nu B/w^2=10$ and $\mu_\nu B/\omega_\nu=5$.}
\label{fig:eigenvalues}
\end{figure}

For computing the coherence, mixedness and trade-off relation, we exploit the 
spectral expansion $\hat{\varrho}(t)=\sum\limits_n\rho_{nn}\vert n\rangle\langle n\vert$, where $\vert n\rangle$ and $\rho_{nn}$ are eigenvectors and eigenvalues of the time-evolved density matrix and $\ln\hat{\varrho}(t)=\sum\limits_n\ln(\rho_{nn})\vert n\rangle\langle n\vert$.
Figure~\ref{fig:eigenvalues} shows the time evolution of the eigenvalues of the density matrix sorted in decreasing order of magnitude. As can be seen, in its eigenbasis, the density matrix has the form of an effective two-level system.

\begin{figure}[ht]
\includegraphics[width=0.45\textwidth]{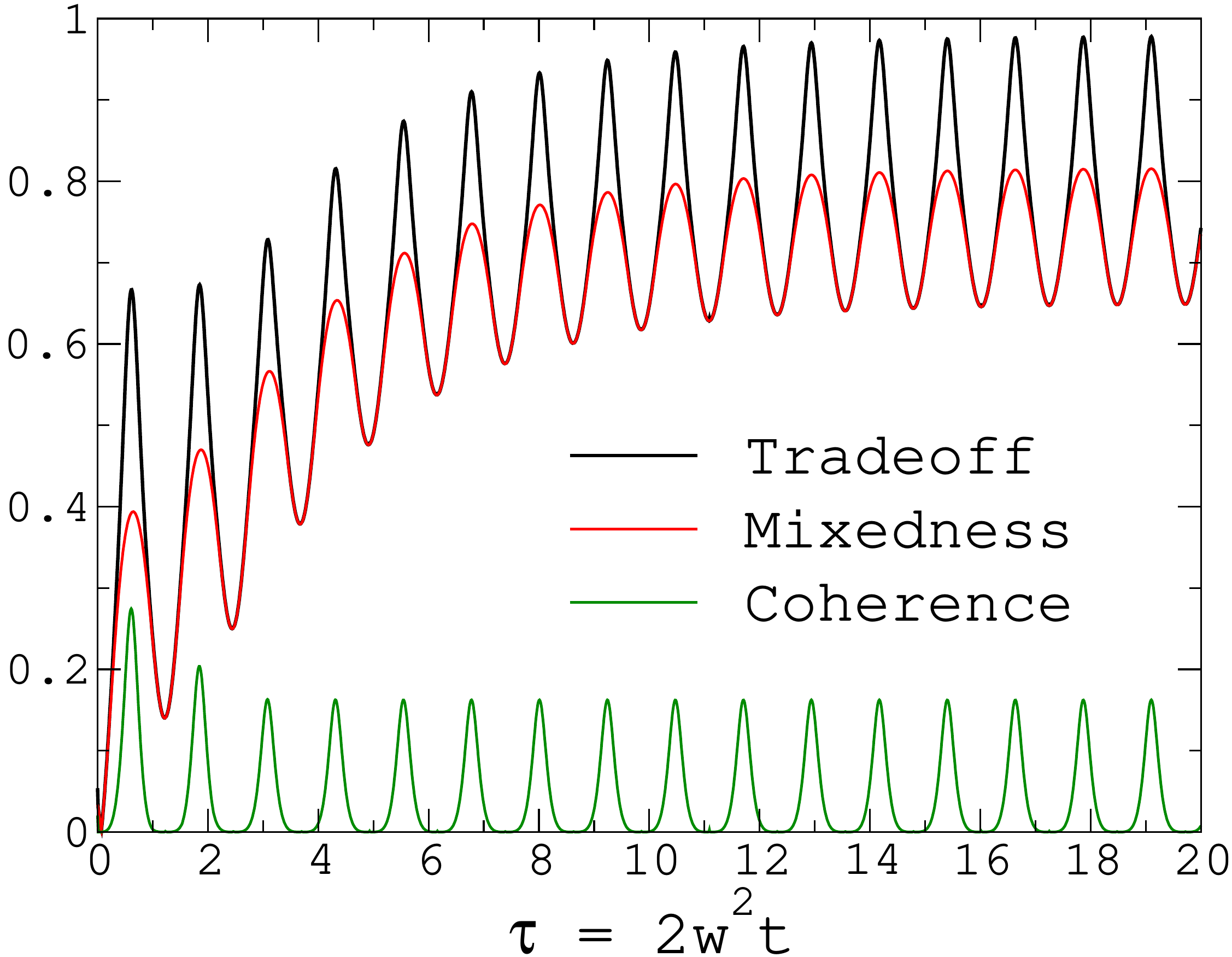}
\caption{The time dependence of coherence, mixedness and their tradeoff when $\beta = \pi/6$, $\homega_B=\mu_\nu B/w^2=10$ and $\mu_\nu B/\omega_\nu=5$.}
\label{fig:tradeoffs}
\end{figure}

The numerical results for the coherence~(\ref{coherence_measure}), mixedness~(\ref{mixedness}) and their trade-off relation~(\ref{the trade off relation}) are shown in Fig.~\ref{fig:tradeoffs}.
The coherence, quantified in terms of the relative entropy, exhibits an interesting behavior with time. Starting from a zero initial value it reaches a maximal value and, after decaying, it goes into a persistent steady-state oscillation regime. 

Thus, we found that after dissipative evolution, the initial spin-polarized state entirely thermalizes, and in the final steady state, the spin-up/down states have the same populations. On the other hand, the flavor states also thermalize. However, the populations of two flavor states do not equate to each other. The initial flavor still dominates in the steady state, and coherence expressed in terms of an entropy measure exhibits persistent oscillations from zero to some constant value which is less than unity.

\section{Conclusions}
Traditionally optical photons and electromagnetic interaction were the primary sources for astronomers to study the distant universe. However, after technological progress achieved during the last few decades, using messengers of other fundamental interactions in the multimessenger astronomy became experimentally feasible \cite{meszaros2019multi}. Exploiting neutrino beams for interstellar communication or other purposes, i.e., navigation \cite{huber2010submarine}, is a  demanding challenge on both theoretical and practical levels.  
Owing to the weak interaction with matter, neutrino beams have the advantage to penetrate the areas where electromagnetic waves are damped.
Due to the spin, flavor, and spin-flavor oscillations, neutrino beams are the essence of not a single but superposition states. Therefore the phenomenon of quantum coherence plays an essential role in the multimessenger astronomy. When the cosmic neutrino beam traverses dissipative interstellar space, the superposition of the different flavor and spin states converts to the mixed state described by the neutrino density matrix.  In the present work, we studied the coupling of the neutrino spin with a random interstellar magnetic field and developed a framework for treating and quantifying the dissipation and coherence effects in neutrino propagation and oscillations. The stochastic field thermalizes the spin state and, due to the spin-flavor channel, impacts the flavor states as well.  We observed that the system never thermalizes to the absolutely mixed state. Trade-off theorem holds, and coherence is preserved in the final steady state. We believe that persistent spin-flavor coherence may play an essential role in the multimessenger astrophysics and neutrino quantum information protocols in the foreseeable future.

\begin{acknowledgments}
We acknowledge financial support from DFG through SFB 762 and SFB TRR227. This work was supported by Shota Rustaveli National Science Foundation of Georgia (SRNSFG) [grant number FR-19-4049]. The work of K.A.K. and A.I.S. is supported by the Russian Foundation for Basic Research under grant no. 20-52-53022-GFEN-A.
\end{acknowledgments}
\bibliographystyle{apsrev4-1}
\bibliography{nems}

\end{document}